\documentclass[letterpaper]{aa501}
\usepackage{natbib}

\newcommand{\BV}{Brunt-V\"ais\"al\"a\,}

\begin{document}

\title{Stratified disks are locally stable}

\author{Jeremy Goodman\inst{1} \and Steven A. Balbus\inst{2}}

\institute{Princeton University Observatory, Princeton, NJ 08544
\and Dept.~of Astronomy, University of Virginia, Charlottesville,
VA 22903}



\authorrunning{Goodman \& Balbus}

\date{Received ; accepted }

\abstract{
Notwithstanding recent claims by Richard \emph{et al.},
there is no linear hydrodynamic instability of axisymmetrically
stable disks in the local limit.  We prove this by means of an
exact stability analysis of an unbounded incompressible
flow having constant stratification and constant shear.
}

\maketitle

\section{Introduction}
In a very recent preprint, \cite{Richard01} claim
to demonstrate nonaxisymmetric local incompressible linear instability of
shearing disks with vertical stratification.  (They also
discuss more general flows that are stratified in both radius
and height).  The growth rate is approximately the geometric
mean of the shear rate ($S$) and \BV frequency ($N$), hence
quite rapid since $|S/\Omega|=3/2$ in keplerian disks, and typically
one expects $N\sim\Omega$ away from the disk midplane.  If true,
this would be very important, as the instability might support turbulence
and angular-momentum transport in disks too weakly ionized for
magnetorotational instabilities \citep{BH98}.

In \S\ref{method}, we derive the exact equation (\ref{spring}) for
time-evolution of spatial Fourier modes in a Boussinesq flow
with a linear shear profile.  In \S\ref{results}, we use this
equation to show that all local disturbances are bounded
at late times, and hence there is no growing local mode. We
also briefly discuss how global instabilities might arise in the
presence of reflecting boundaries.

\section{Local Boussinesq model}\label{method}

Using a local WKB analysis, \cite{Richard01} assert that local Fourier
modes with wavenumber $\vec{k}=(k_r,k_\theta,k_z)$
grow $\propto\exp(\Gamma t)$ 
\footnote{The term $k_r^2 N^2$ in the denominator is missing in the 
expression quoted in \cite{Richard01}.  This appears to be
a typographical error.  Whether the term is present or not
does not affect our critique.}
\begin{equation}\label{Rgrowth}
\Gamma^2 = -\frac{2k_\theta^2 N^2 \Omega S}{k_\theta^2(N^2+2\Omega S)
+ k_r^2 N^2 + k_z^2\kappa^2}.
\end{equation}
As usual, $r\Omega(r)\vec{e}_\theta$ is the background orbital velocity,
$S=d\Omega/d\ln r<0$ is the radial shear, and
\begin{eqnarray*}
\kappa^2 &=& r^{-3}\frac{d(r^4\Omega^2)}{dr}~=2\Omega(2\Omega+S),\\
N^2      &=& -\frac{1}{\gamma\rho}\frac{\partial P}{\partial z}\,
\frac{\partial }{\partial z}\ln\left(\frac{P}{\rho^\gamma}\right)
\end{eqnarray*}
are the squares of the epicyclic and \BV frequencies.  Since it
is assumed that $N^2>0$ and $\kappa^2>0$, the flow would be
stable against both convection and centrifugal instability if
there were no shear.  Yet according to eq.~(\ref{Rgrowth}),
there is a constant growth rate $\omega\sim\sqrt{-2\Omega S}$
in the limit of strong stratification $N^2\to\infty$ provided
only that $d(\Omega^2)/dr<0$ (as is normally the case in astrophysical
disks).

In the extreme local limit, one considers disturbances on scales
very small compared to the distance to the nearest boundary of
the flow.  Also, provided that $\Omega$ and $N^2$ are smooth
functions, they vary little on the small lengthscale.
It is therefore reasonable to consider an unbounded flow in
which all of the frequencies $\Omega,N,S,\kappa$ are constant
in space and time.  Variations in density can be ignored except
in the buoyancy term (Boussinesq approximation).  The governing
equations for this flow in a corotating frame are
\begin{eqnarray}\label{model}
0&=& \partial_t\vec{v} +\vec{v}\cdot\vec{\nabla}\vec{v} 
+2\Omega\vec{e}_z\times\vec{v} +\vec{\nabla}\psi
+N^2 L\,\vartheta\vec{e}_z,\nonumber\\
\frac{v_z}{L}&=& \partial_t\vartheta +\vec{v}\cdot\vec{\nabla}\vartheta,
\nonumber\\
0&=& \vec{\nabla\cdot}\vec{v}.
\end{eqnarray}
The quantity $\vartheta$ represents the fractional density difference
between a displaced fluid element and the background (which the
Boussinesq approximation considers to be infinitesimal). $L$ is a constant
scale height for the entropy, so that $N^2=g/L$ if $g$ is
the vertical gravitational acceleration.  $\psi$ is
a scalar potential for accelerations due to gravity and pressure.
The background state is
\begin{equation}\label{background}
\vec{v}_0=Sx\vec{e}_y,\quad \vartheta_0=0,\qquad\psi_0=\Omega S x^2.
\end{equation}
The coordinates $r,\theta,z$ have been replaced by local Cartesians $x,y,z$.
Were \cite{Richard01}'s analysis correct, 
their growth rate (\ref{Rgrowth}) would apply to this model.

At any given time ($t$), linearized disturbances of this flow can be
be decomposed into spatial Fourier components,
\begin{equation}\label{pform}
\vec{v}_1 = \epsilon\vec{V} e^{i\vec{k}\cdot\vec{r}},\quad
\vartheta_1 = \epsilon\Theta e^{i\vec{k}\cdot\vec{r}},\quad
\psi_1 = \epsilon\Psi e^{i\vec{k}\cdot\vec{r}},
\end{equation}
where $\epsilon\ll 1$ is a formal small parameter.
Following a method attributed to \cite{Kelvin},
(\ref{pform}) is a solution to the linearized equations of motion
for \emph{all} time
if $\vec{V},\,\Theta,\,\Psi$ \emph{and} $\vec{k}$ are appropriate
functions of $t$ [but independent of position $\vec{r}=(x,y,z)$].
Note that this means that the time dependence of the flow attributes
will not, in general, be a simple exponential function, in contrast
to the \cite{Richard01} assumption.  
Substituting into the first of eqs.~(\ref{model}) and collecting terms
of first order in $\epsilon$,
\begin{eqnarray*}
0&=& \left(\frac{d\vec{k}}{dt}\cdot\vec{r}\,+Sxk_y\right)i\vec{V} \\
&+&\left(\frac{d\vec{V}}{dt}
+SV_x\vec{e}_y+2\Omega\vec{e}_z\times\vec{V}+i\vec{k}\Psi
+N^2L\Theta\vec{e}_z\right).
\end{eqnarray*}
The coefficients of each coordinate
$xyz$ must vanish separately, whence
\begin{equation}\label{kevol}
\frac{dk_x}{dt}=-Sk_y,\quad\frac{dk_y}{dt}=\frac{dk_z}{dt}=0,
\end{equation}
and
\begin{equation}\label{Vevol}
\frac{d\vec{V}}{dt}+SV_x\vec{e}_y+2\Omega\vec{e}_z\times\vec{V}+i\vec{k}\Psi
+N^2L\Theta\vec{e}_z =0.
\end{equation}
By similar steps,
\begin{equation}\label{Tevol}
\frac{d\Theta}{dt}=\frac{V_z}{L}
\end{equation}
and
\begin{equation}\label{divv}
\vec{k}\cdot\vec{V}=0~\Leftrightarrow~ V_z=-\frac{k_xV_x+k_yV_y}{k_z}\,.
\end{equation}
Combining the time derivative of (\ref{divv}) with eq.~(\ref{Vevol})
yields
\begin{equation}\label{Psol}
\Psi=\frac{i}{k^2}\left[2(S+\Omega)k_yV_x-2\Omega k_xV_y+N^2Lk_z\Theta
\right].
\end{equation}
Note the abbreviations
\begin{equation}\label{abbrev}
k^2=k_x^2+k_y^2+k_z^2,\qquad k_\perp^2=k_x^2+k_y^2=k^2-k_z^2.
\end{equation}

Since $V_z$ can be eliminated \emph{via} eq.~(\ref{divv}),
the system (\ref{Vevol})-(\ref{Tevol}) is third order in time.\footnote{
Yet \cite{Richard01}'s dispersion relation is fourth-order in
frequency.  At $k_y=0$, two roots vanish, and
the claim is made that these develop into $\pm\Gamma$ as given by
(\ref{Rgrowth}) when $k_y\ne 0$.  Non-vortical perturbations, however,
are restricted to obey a \emph{second} order system. 
Moreover, if time dependences were in fact
exponential, then a nonzero value of the vorticity constant in
(\ref{pvort}) would correspond to a zero eigenfrequency at \emph{any} $k_y$.}
However, it has a first integral,
\begin{equation}\label{pvort}
k_x U_y -k_y U_x -(2\Omega+S)k_zL\Theta=\mbox{constant}
\end{equation}
representing conservation of potential vorticity.
For growing modes, the constant would be
asymptotically negligible.  So we set it to zero.
Eliminating $V_z$ between
eqs.~(\ref{Tevol}) and (\ref{divv}):
\begin{equation}\label{conver}
k_xU_x+k_yU_y=-k_z L\frac{d\Theta}{dt}.
\end{equation}
We use eqs.~(\ref{pvort}) \& (\ref{conver}) to eliminate
$U_x$ and $U_y$ from eq.~(\ref{Vevol}) in favor of $\Theta$ \& $\dot\Theta$:
\begin{eqnarray}\label{ode0}
0&=&\frac{d^2\Theta}{dt^2} \,+\left(\frac{2Sk_xk_yk_z^2}{k^2k_\perp^2}\right)
\frac{d\Theta}{dt} \nonumber\\
&+&\left[N^2\frac{k_\perp^2}{k^2}+\kappa^2\frac{k_z^2}{k^2}
+2S(2\Omega+S)\frac{k_y^2k_z^2}{k^2k_\perp^2}\right]\Theta.
\end{eqnarray}
(Note that eq.~(\ref{ode0}) also describes the evolution
of vertical Lagrangian displacements, a consequence of eq.~(\ref{Tevol}).)
The variable
\[ \Phi=\frac{k}{k_\perp}\Theta
\]
obeys a slightly tidier equation:
\begin{eqnarray}
0 &=&\frac{d^2\Phi}{dt^2} \,+\omega^2(t)\Phi ,\nonumber\\
\label{spring}
\omega^2(t)&=& N^2\frac{k_\perp^2}{k^2}+\kappa^2\frac{k_z^2}{k^2}
\nonumber\\
&+&\frac{k_y^2k_z^2}{k^2k_\perp^2}\left(4\Omega S -\frac{k_x^2}{k^2}S^2
+\frac{3k_y^2}{k_\perp^2}S^2\right).
\end{eqnarray}
Together with the solution of eq.~(\ref{kevol}),
\begin{equation}\label{ksol}
k_x(t)=k_x(0)-2Stk_y,\qquad k_y,k_z=\mbox{constants},
\end{equation}
eq.~(\ref{spring}) governs the evolution
of a Fourier mode for the density perturbation.
The equation is exact within the context of the model defined
by eqs.~(\ref{model})-(\ref{background}).  In fact, it is
not even restricted to small amplitudes, because the terms
quadratic in $\epsilon$ vanish when eq.~(\ref{Psol}) is substituted
into eq.~(\ref{model}) as a consequence of eq.~(\ref{divv}).

\section{Discussion}\label{results}

Clearly, eq.~(\ref{Rgrowth}) is incompatible with eq.~(\ref{spring}).
When $k_z = 0$, for example, eq.~(\ref{Rgrowth})
predicts a finite growth rate.  The exact solution of eq.~(\ref{spring})
corresponds to vertical oscillations at the \BV frequency, an
elementary result (consider purely vertical displacements) that may
be obtained directly from the $z$ component of equation
(\ref{Vevol}).

More generally,
as $St\to\pm\infty$, $k_\perp^2\approx k^2\approx k_x^2~\gg\, k_y^2,k_z^2$,
so that $\omega^2(t)\to N^2$ and $\Phi\propto\exp(\pm iNt)$.
Hence, contrary to the result claimed by \cite{Richard01} 
[eq.~(\ref{Rgrowth})],
{\it there is no exponentially growing linear mode}.\footnote{
This assumes that the disk is stably stratified.  On the
contrary, when $N^2<0$, an exponentially growing convective disturbance
persists as $k_x\to\infty$ if, as here,
viscosity and thermal conduction are ignored \citep[\emph{e.g., }][]{RG92}.}

An approximate dispersion relation can be read off from eq.~(\ref{spring})
in the limit $k_x\gg k_y,k_z$ (\emph{i.e.} at late times). The corresponding
radial group velocity is
\begin{equation}\label{vgroup}
\frac{\partial\omega}{\partial k_x}\approx \mp \frac{S(N^2-\kappa^2)}{N}\,
\frac{k_z^2k_y}{k_x^3}
\end{equation}
This tends to zero as $t^{-3}$.
It can be shown that a radially localized wavepacket, obtained by superposing
Fourier modes with a range of $k_x(0)$, asymptotically approaches resonant
radii where the frequency of the disturbance $=\pm N$ in a frame
corotating with the local background.  If the disturbance is excited
with azimuthal wavenumber $k_y$
by a force corotating steadily at radius $r_0$,
the resonances lie approximately at $r_0\,\pm N/Sk_y$.
In order that our local analysis apply to radially bounded flows,
it is necessary that these resonant radii lie inside the fluid.
Otherwise the wavepacket may reflect from the boundary, suffering
a change in the sign of $k_x$, and offering the possibility of repeated
passages through the ``swing amplifier'' at small $k_x/k_y$ where $\omega^2$
can be briefly negative.\footnote{See \cite{Toomre81} and \cite{GT78} for
particularly clear discussions of the swing-amplifier in the context of
two-dimensional compressible disks.}

The instability discussed by by \cite{Molemaker01} for stratified,
centrifugally-stable Couette flow, and cited by \cite{Richard01} as
support for their claims of instability, is an intrinsically global
phenomenon, and has no local counterpart.  \cite{Molemaker01} analyzed
their instability in terms of trapped Kelvin wave edge modes; it is
also possible that the resonant interaction described above may be
involved in destabilizing some of the modes.  In either
case, the mechanism is certainly not reducible to exponentially-growing
local \BV oscillations.  A full analysis of the global problem is
beyond the scope of this note.

\acknowledgements
JG is supported by NASA grant NAG5-8385; SAB by NASA grants NAG 5-7500
and NAG-106555.

{}

\end{document}